# Optimization and Benchmarking of Monolithically Stackable Gain Cell Memory for Last-Level Cache

Faaiq Waqar, Jungyoun Kwak, Junmo Lee, Omkar Phadke, Minji Shon, Mohammadhosein Gholamrezaei, Kevin Skadron, Shimeng Yu

*Abstract—* **The Last Level Cache (LLC) is the processor's critical bridge between on-chip and off-chip memory levels - optimized for high density, high bandwidth, and low operation energy. To date, high-density (HD) SRAM has been the conventional device of choice; however, with the slowing of transistor scaling, as reflected in the industry's almost identical HD SRAM cell size from 5 nm to 3 nm, alternative solutions such as 3D stacking with advanced packaging such as hybrid bonding are pursued (as demonstrated in AMD's V-cache). Escalating data demands necessitate ultra-large on-chip caches to decrease costly off-chip memory movement, pushing the exploration of device technology towards monolithic 3D (M3D) integration where transistors can be stacked in the back-end-of-line (BEOL) at the interconnect level. M3D integration requires fabrication techniques compatible with a low thermal budget (<400°C). Among promising BEOL device candidates are amorphous oxide semiconductor (AOS) transistors, particularly desirable for their ultra-low leakage (<fA/μm), enabling persistent data retention (>seconds) when used in a gain-cell configuration. This paper examines device, circuit, and system-level tradeoffs when optimizing BEOL-compatible AOS-based 2-transistor gain cell (2T-GC) for LLC. A cache early-exploration tool, NS-Cache, is developed to model caches in advanced 7 & 3 nm nodes and is integrated with the Gem5 simulator to systematically benchmark the impact of the newfound density/performance when compared to HD-SRAM, MRAM, and 1T1C eDRAM alternatives for LLC.**

## I. INTRODUCTION

The ever-increasing demand for computing power, driven by innovations in artificial intelligence (AI), machine learning (ML), and scientific computing, is expected to push high-performance systems to the zettascale ($10^{21}$ operations/s) [1]. This rise in computational performance is met with an increasing demand for memory bandwidth and capacity at the last-level cache (LLC), brought about by the greater bandwidth capabilities of off-chip memory (HBM), system architectures and communication paradigms

This manuscript is a preprint submitted to arXiv. This work was supported by PRISM, a center of the SRC/DARPA JUMP 2.0 Program. The authors thank M. U. Karim of Samsung for guiding discussions on experiments conducted in this study.

Faaiq Waqar, Jungyoun Kwak, Junmo Lee, Omkar Phadke, Minji Shon, and Shimeng Yu and with the School of Electrical and Computer Engineering, Georgia Institute of Technology, Atlanta, GA, USA (email: {faaiq.waqar, jkwak38, junmolee, omkarphadke, mshon6}@gatech.edu, shimeng.yu@gatech.edu).

Mohammadhosein Gholamrezaei and Kevin Skadron are with the Department of Computer Science, University of Virginia, Charlottesville, VA, USA (uab9qt@virginia.edu; skadron@virginia.edu).

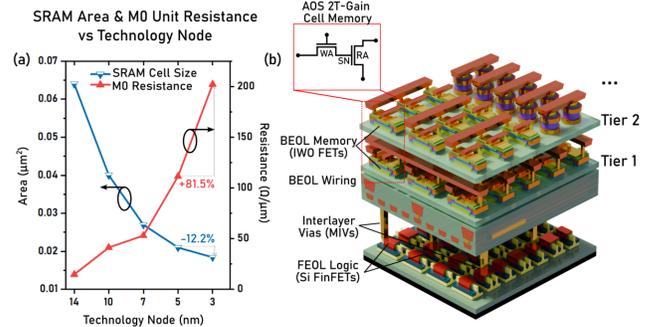

Fig. 1. (a) Scaling of the high density (HD) bit-cell and lower metallization resistance in FinFET generation technology nodes, derived from IRDS and foundry reports [19]. (b) low-temperature monolithic 3D integration of AOS transistors in the BEOL (above CMOS peripherals) reveals unprecedented opportunity for denser cache memory.

that exacerbate data traffic at the shared LLC (e.g., CXL [2]), and the increased prevalence and demands of data-intensive workloads, such as those used for AI/ML and scientific computing applications. To improve the miss rate of the LLC and therefore limit costly (i.e., energy, latency) off-chip data movement, it is desirable to construct ultra-high-capacity LLCs to keep pace; however, leading-edge FinFET-generation high-density (HD) SRAM, the conventional LLC memory of choice, scales slowly in 7/5/3 nm nodes compared to prior generations (Fig. 1a). As an example, TSMC's minimum reported HD SRAM bit cell only shrank from 0.021 μm$^2$ in 5 nm (N5) to 0.0199 μm$^2$ in 3 nm (N3B) processes [3]. Therefore, it has been projected that as much as ~50-70% of silicon (Si) real estate may be occupied by SRAM in modern processors [4]. Meanwhile, interconnect parasitics have compounded due to reduced pitch and high resistive diffusion barrier interaction, which increases grain boundary and surface scattering, thus leading to longer RC delays. To combat these challenges, the adoption of the double-word line (reducing WL resistance) has been suggested [5]. Another approach is the advanced packaging of 3D die-stacked SRAM caches, which AMD has demonstrated commercially in V-Cache [6]. However, hybrid bonding is (currently) a costly process, and the pitch of bonding pad connections (~a few μm) limits the die-to-die bandwidth.

To maximize the potential of monolithically integrated caches (i.e., without advanced packaging techniques), several alternative cache memory devices have garnered notable attention for their desirable footprint, absence of direct leakage paths, and short access latency (Table I). Logic-compatible embedded DRAM (1T1C eDRAM) has seen implementation in commercial L3 caches, including IBM's POWER 8/9 [7] and Intel's Skylake processors [8], with a footprint 4× smaller than SRAM in the 14 nm technology node. However, optimizations used in commodity DRAM, such as recessed/saddled channel and pillar capacitor processes [9], cannot be used with logic processes. This leads to challenges in scaling eDRAM technology and short μs-level retention, which limits bandwidth (BW) and increases energy consumption due to frequent refresh operations. Spin-transfer-torque magnetic random-access memory (STT-MRAM) has garnered popularity as an LLC candidate in the past few years for its long retention, low read energy, and high cycling endurance [10]. However, MRAM's high write current/latency and small sense margin limit the scaling potential of the bit cell (despite miniaturized magnetic tunnel junction (MTJ)) and LLC read/write bandwidth. For these reasons, STT-MRAM's commercial target refocused onto embedded Flash replacement for automotive microcontrollers rather than LLC in a processor.

Materials advances within the device research community are actively enabling the fabrication of transistors in the back-end-of-line (BEOL) without damaging the underlying front-end-of-line (FEOL) active Si devices [12] by utilizing amorphous oxide semiconductor (AOS) channel materials (Section 2C). Transistors with channels made from n-type AOS materials demonstrate adequate mobility of ~20 cm$^2$/V-s, low leakage (<fA/μm) [11], and can be fabricated at low temperatures (<300°C). In a two-transistor gain-cell (2T-GC) configuration, where the storage capacitor in an eDRAM cell is swapped for a second access transistor and read/write paths are bifurcated (leading to a non-destructive readout), AOS devices are advantageous, as their low leakage enables long retention (> seconds). The prospect of M3D integrated 2T-GC caches is appealing because the majority of LLC area is consumed by memory (e.g., with area efficiency 65-83%), meaning that the potential for area reduction using AOS devices far surpasses that of alternatives with a FEOL presence if the 2T-GCs are stacked above CMOS peripherals (Fig. 1b).

Extensive studies at the device level [11] and subarray-level [12]-[13] have been conducted on 2T-GCs with AOS devices based on In$_2$O$_3$ doped with Ga & Zn (IGZO) [14], Sn (ITO) [13], and W (IWO) [11]. However, the architectural modeling of AOS LLC integration at scale in which the challenges of high capacity (parasitic accumulation, bandwidth limitations, etc.), connectivity (subarray, mat, and peripheral organization), and performance on variable transaction patterns that are characteristic of cache memory (as opposed to predictable FIFO-like flows) has not been performed. The challenge of doing so arises from a combination of limited device data (primarily reported at top-tier device conferences such as IEDM/VLSI) and a lack of early-exploration modeling tools that support advanced technologies beyond the 22 nm bulk transistor regime [15]. The work presented in this paper aims to remedy both challenges by meticulously modeling AOS devices in physics-based Technology-CAD (TCAD) simulation, developing RC-based cache-exploration tools to explore the design space of caches in FinFET and nanosheet nodes (14 − 1 nm), and modifying/interfacing with an architectural simulator (Gem5) to run workloads on the proposed LLC designs (Fig. 2).

Herein, we perform a pioneering systematic examination of AOS 2T-GC memory optimized for ultra-high-capacity LLCs towards 100 MB and above. Through a comprehensive modeling flow, we elucidate guidelines that can be used cross-spectrum: from device engineers looking to optimize AOS transistors for further performance gain to system designers understanding how to maximize the density benefits of AOS-based M3D designs. The contributions of this work are:

- Using Sentaurus TCAD, we design and study IWO 2T-GCs in scaled double-gate (DG) and channel-all-around (CAA) vertical structures and study tradeoffs in speed, density, and retention when modulating threshold voltage ($V_{th}$), defect (i.e., oxygen vacancy) profile, gate to source/drain overlap, and supply voltages asymmetrically.

- We introduce NS-Cache, an exploratory power-performance-area (PPA) RAM/Cache analysis tool for leading-edge nodes and M3D designs. NS-Cache interfaces with Gem5's Ruby protocol system to provide accurate timing for exploring caches with refresh operations and various access modes.

- We propose and analyze a **T**ag **A**rrays **U**nder Data (TAU) M3D integration scheme to offset the overhead of (SRAM) tag memories and reduce hit latency in ultra-large AOS 2T-GC caches.

- We benchmark SRAM, 1T1C eDRAM, STT-MRAM, and AOS 2T-GCs at 7 nm node to compare viable cache candidates using NS-Cache and Gem5 at iso-area/capacity. Furthermore, we pit a bandwidth-optimized AOS 2T-GC cache at the 7 nm node against a comprehensive AMD's Zen3 3D V-Cache model.

- We compare SRAM and vertical CAA AOS transistors at the macro-level in the 3 nm node.

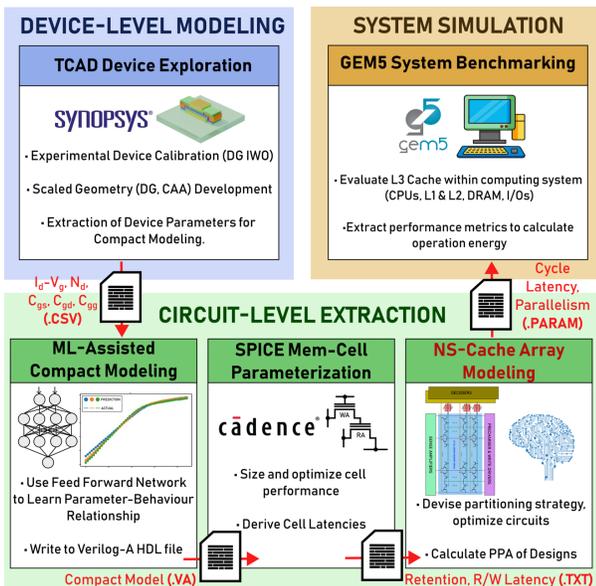

Fig. 2. LLC Modeling flow for device, circuit, and architectural PPA exploration of AOS 2T-GC LLCs. NS-Cache is open-source on GitHub.

## II. Background and Prior Work

### A. Cache Organization and Early Exploration Frameworks

CACTI [16], first developed by Hewlett Packard (HP) in 1993, is a well-known SRAM cache modeling tool, although it has several extensions used to model die-stacked 3D DRAM, off-chip I/O, etc., developed over six major revisions. CACTI's modeling of bank-level power performance and area (PPA) uses the logical effort methodology, which balances drive strength across logical paths to minimize propagation delay, using a combination of technology (transistor), fanout, and RC (Horowitz, Elmore) models. NVSim [17] was constructed out of CACTI using similar design principles but appended support for emerging non-volatile memory models such as MRAM, resistive random-access memory (RRAM), and phase change memory (PCM), as well as bus routing methodology for caches employing snooping. DESTINY [15] added support for 3D eDRAM and NVM cache designs to the NVSim base and validated models for peripheral circuits based on experimentally demonstrated chips. However, DESTINY does not support transistor models beyond 22 nm bulk technology, leaving a gap in cache/RAM modeling in FinFET generation nodes and beyond. NeuroSim [18], initially developed for compute-in-memory architectural analysis, integrates predictive FinFET and nanosheet transistor exploration down to the future 5Å CFET node through rigorous device-level characterization derived from industry and IRDS projections [19]. NeuroSim V1.4 and Destiny share NVSim as a common ancestor, which allows us to synthesize a tool that combines and expands upon both tools: NS-Cache. NS-Cache is built to provide the community with an open-source exploration of advanced novel cache designs in leading-edge nodes and the design of M3D caches using fully BEOL-compatible memories (available at github.com/neurosim/NS-Cache).

In CACTI, NVSim, DESTINY, and NS-Cache, a cache's hierarchical organization is divided into banks, subarrays, and mats from the top down (Fig. 3). The bank represents the top-level independent structure, containing a grid of subarrays interconnected by an H-tree or bus-like routing scheme used to facilitate the movement of data. In some commercial caches, such as those used in AMD's Zen architecture, sets of banks may be partitioned into a higher-level independent structure called slices. During an L3 transaction, an address incoming from the memory management unit (MMU) or L2 is first routed through I/O logic (❶) and passed to control circuitry (❷), which encodes and routes control signals over the interconnect (❸) to a set of concurrently operating subarrays. A subset of the number of rows ($N_{SR}$) and columns ($N_{SC}$) of subarrays are activated during a transaction, deemed "active," and are described by active rows ($N_{ASR}$) and columns ($N_{ASC}$), meaning the total number of active subarrays is $N_{ASC} \times N_{ASR}$. For example, if $N_{ASR} = N_{ASC} = 2$, four subarrays are activated during each transaction, indicating that groups of four subarrays may be operated independently (assuming controller support). The number of bits routed to each subarray differs in quantity based on bank type and access mode, which are discussed in further detail in Section 4C and Table II. The number of routed bits is a function of associativity ($A$), block width ($W_{Block}$) and the number of blocks ($N_{Block}$). In data banks, $W_{Block}$ is defined by the word width and parallelism between subarrays to a single data-retrieval operation. Each subarray has a uniform number of blocks, determined by $N_{Block}/N_{Subarray}$. In tag banks, the physical addresses of entries are stored in the corresponding data entries without offset/index bits, and dirty + valid bits are included for coherency. A pre-decoder (❹) breaks down the address at the subarray level to activate mats over the subarray interconnect (❺). Mats within a subarray may act as distributed members, each fulfilling cache-line operations in parallel composed of broken-down blocks to deliver a complete block-width subarray operation, depending on their activation ($N_{AMR}$, $N_{ASC}$). To maximize parallelism and offset the latency of routing ($>t_{Subarray}$), subarrays may be pipelined, in which case the bandwidth of the cache is bound by the subarray delay, making cell access time a critical parameter to optimize. Mat components are discussed in further detail in Section 4. We note that the nomenclature for subarrays and mats in NS-Cache is swapped from DESTINY/NVSim to maintain consistency with existing literature on cache organization.

### B. Cache Memory Device Candidates

The LLC is optimized for density, bandwidth, and operation energy in high-performance processors. Naturally, HD SRAM, with minimal sizing in pull-down (PD), pull-up (PU), and pass-

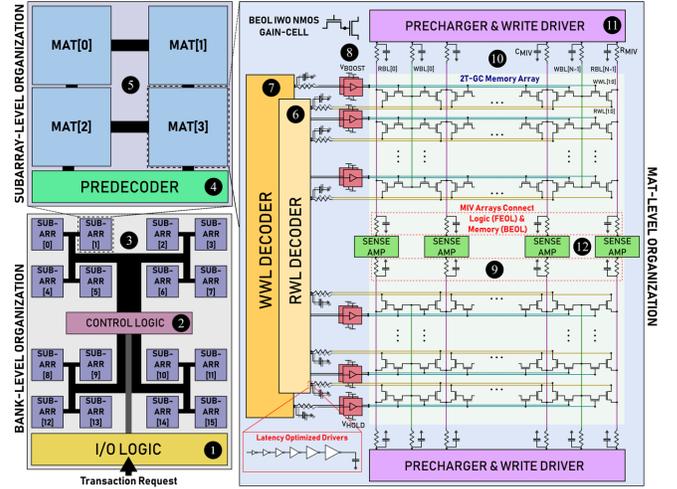

Fig. 3. Organization of an AOS 2T-GC cache from the top (bank) down. Peripheral circuits in the FEOL dictate density at the mat level; therefore, footprint reductions are passed from the bottom up in M3D design.

### TABLE I
#### Comparison of Cache Memory Device Candidates

| FOM / Cell | HD 6T SRAM | 1T1C eDRAM | STT-MRAM | AOS 2T-GC* |
|---|---|---|---|---|
| Process | FEOL (CMOS) | FEOL (Access) + FEOL (DTC) or BEOL (STC)** | FEOL (Access) + BEOL (MTJ) | BEOL (Read/Write Access) |
| Retention | Infinite*** | Short | Long | Medium-Long |
| Leakage | High (>10 pW) | Low | Low | Low |
| Access Latency | Low (<100 ps) | Low | High (>5 ns) | Low-Medium |
| Access Energy | Low | Low | High | Low |
| Area (F=FP**) | 25-33 $F^2$ | 8-13 $F^2$ | 11-16 $F^2$ | 14-18 $F^2$ (BEOL) |
| Attributes | (+) Quick Access (-) Large & Leaky | (+) Compact (-) Poor Retention | (+) Persistent (-) Write Access Energy/Latency | (+) 3D Integration Potential (-) In Early R&D |

\* Reported metrics collated from experimental results and simulation data collected in this work
\*\* DTC = Deep Trench Capacitor, STC = Stacked Capacitor, FP = Fin Pitch;  \*\*\* Assuming the supply is not shut off

gate (PG) transistors, is a strong choice. Push rules and folding are employed by foundries to reduce bit cell size further [24]. However, subthreshold leakage in the inverter pair is responsible for the sizeable static power consumption that limits on-chip energy efficiency in SRAM caches, which cannot be mitigated using power gating due to the volatility of the cell. As an illustration of this, we find that in an 8 kB SRAM mat in 7 nm, ~86% of the total leakage power is consumed by memory cells in standby mode. Beyond the stacked nanosheet era (1 nm), complementary FET (CFET) (e.g., n-type of nanosheet stacked on top of p-type) for the Angstrom era have been proposed to reinvigorate SRAM bit-cell scaling; however, stagnated footprint reduction is expected to repeat subsequent to the CFET transition at ~0.0105 $\mu m^2$ [19].

STT-MRAM is a 2-terminal memory switched between low and high resistance states by modulating the magnetization in the free ferromagnetic layer between parallel and anti-parallel to that of the pinned synthetic anti-ferromagnetic layer. Prior works have explored the non-volatility and write latency tradeoff in STT-MRAM to improve its viability as a cache memory [10]. STT-MRAM's high-current (>100 $\mu A$) write operation requires a large access transistor (2-3 fins in FinFET) built into the FEOL; thus, STT-MRAM's bit cell size reduction factor is limited to ~2-3× compared to an HD SRAM cell. Additionally, a low tunneling magneto-resistance ratio (TMR) results in a small sense margin, deteriorating read speed, and requiring sizeable current sense amplifiers (CSA) and reference cell strings [21]. Cache-suitable STT-MRAM is experimentally demonstrated down to the 14 nm node by major foundries [22], with scaling projections towards the 5 nm node [23].

1T1C eDRAM is schematically no different from its main-memory off-chip counterpart; a storage node (SN) capacitor stores charge to represent a bit. However, logic compatibility produces two challenges. The SN has lower capacitance since a high aspect ratio cylindrical structure cannot be used, leading to a degraded sense margin. Second, access transistor innovations such as gate recessing/saddling cannot be employed. This leads to weak retention from the gate-induced drain (GIDL) and subthreshold leakages ($I_{leak}$), orders of magnitude lower than the

JEDEC standard for tail bits in commodity DRAM (32 ms at 85°C). Intel has demonstrated a second-generation eDRAM with 300 $\mu s$ retention in a 14 nm platform at 95°C [8]. However, the industry has not reported further scaling of 1T1C eDRAM. Using a calibrated LP-NMOS SPICE model, we project that 1T1C eDRAM's retention would be 170 $\mu s$ if scaled to the 7 nm technology node.

The 2T-GC trades the dedicated SN capacitor in a 1T1C eDRAM cell for a read transistor. The drain/source of the write transistor is connected to the gate of the read transistor; thus, the gate capacitor of the read transistor becomes the SN. Read (RA) and write access (WA) are disjoint, giving a non-destructive readout using the transconductance of the read transistor when a differential voltage is applied across the read bitline (RBL) and read wordline (RWL). Si 2T-GCs may be constructed asymmetrically [24] with both PMOS and NMOS. When constructed from AOS transistors, only n-type devices are employed, as p-type oxide transistors have poor mobility, leading to extended access latency [25]. In Si, the 2T-GC suffers similar retention challenges to eDRAM, worse given the lack of a dedicated SN capacitor; however, AOS 2T-GCs can have orders of magnitude higher retention (e.g., seconds at 85°C) thanks to their exceptionally low leakage due to the wide band gap of AOS materials [11]. However, the relatively lower mobility relative to FEOL Si transistors may sacrifice its access speed by 1-2 orders of magnitude (a tradeoff discussed further in Section 3). It should be noted that the n-type only AOS 2T-GC suffers from a capacitive coupling issue that fluctuates the SN voltage when a rise/fall event occurs at WL or BL. Thus, advanced geometric techniques such as the split-gate design have been proposed to mitigate such adverse effects [42].

## C. Monolithic 3D Integration

Whereas heterogeneous 3D (H3D) approaches seek to increase integration density by stacking discrete dies/chiplets using advanced packaging, M3D integration pursues increased transistor density by building active tiers above Si devices where the BEOL interconnect resides. It is noted that the two (H3D & M3D) are not developed in mutual exclusivity; in the long term, monolithic, stacked, and assembled systems aim to maximize both the utility on and between dies [28]. To prevent the degradation of underlying CMOS, BEOL-compatible devices must be constructed at temperatures below 400°C [6]. Conventional Si-CMOS technology cannot be monolithically stacked due to the high-temperature annealing process step (~1000°C). One approach to circumvent this is to use targeted device annealing using rapid thermal processing [30] or laser beams when building BEOL devices. Alternatively, one may substitute existing semiconductors with ones that can be processed at low temperatures, such as 2D transition metal-dichalcogenides (2D-MDCs) [27], carbon nanotubes (CNTs) [28], or amorphous oxide semiconductors (AOS) [25] as the channel material of BEOL transistors. An advantage of AOS materials is the decades of development backing their adoption as thin-film transistors commercially used in display technology and flexible electronics, yielding better processing uniformity and scalability for high-yield manufacturing, a problem especially prominent in 2D-MDCs and CNTs. BEOL devices grown in situ are connected with lower-level interconnects using monolithic interlayer vias (MIVs). These MIVs (diameter ~30-50 nm) offer 24× higher I/O density than their H3D counterparts through silicon vias (TSVs), allowing for greater bandwidth than stacked chiplets per unit area [29].

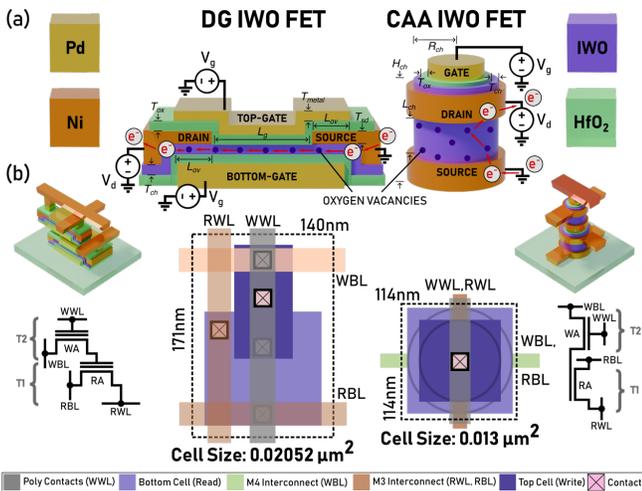

Fig. 4. (a) Structure, materials, and transport of n-type amorphous W-doped $In_2O_3$ channel transistors. (b) Multi-tier AOS 2T-GC schematics and layout under 7 nm (DG) and 3 nm (CAA) design rules.

## III. DEVICE DESIGN AND TRADEOFFS

The structure and electron transport mechanisms in double-gate (DG) and channel-all-around (CAA) AOS transistors with W-doped In$_2$O$_3$ (IWO) channel materials are shown in Fig. 4. The DG transistor structure is a hybridization of coplanar top and bottom gated nanosheet structures that uses both gate electrodes in tandem to improve on-state current and subthreshold swing (SS). Conversely, the CAA structure is optimized for density, built by etching a hole into a metal-insulator-metal (MIM) capacitor, then depositing channel, oxide, and gate materials to form a vertical transistor where channel width is tied to circumference [14]. The gate electrode(s) modulate the current flow across the source/drain contacts through the capacitive accumulation of carriers to form an accumulation layer that acts as the channel [29]. The transport of electrons through the channel is primarily facilitated by donor-like oxygen vacancies ($N_D$). Here, we focus on a particular channel IWO among various AOS materials for the analysis owing to the mitigated $V_{th}$ shift (<100 mV at 85ºC) under the bias-temperature instability (BTI), a result of the W doping effect into the In$_2$O$_3$ material [43].

An intrinsic tradeoff exists between access speed and retention when choosing a threshold voltage ($V_{th}$) for the write access device, as illustrated in:

$$t_{r/a} = \int_{v_1}^{v_2} \frac{c_{sn}(v_{sn})}{I_{leak/on}(v_{hold/boost}, v_{sn})} \, \partial v_{sn}, \; \log_{10}(\frac{I_{on}}{I_{leak}}) \approx \frac{\Delta V_g}{SS} \quad (1)$$

Where $t_{r/a}$ is the retention/access time, $c_{sn}$ is the storage node capacitance, and $SS$ is the subthreshold slope. In retention, {$v1$, $v2$} represents the voltage value of a "1" and the minimum voltage at the SN to give an adequate sense margin, and in access, they represent voltage values of "0" to "1". Because $\Delta V_g$ is limited by voltage swing, and $SS$ is limited by the Boltzmann limit (60 mV/decade at 25°C), the on/off current ratio is ultimately confined. This tradeoff is further pronounced by AOS devices' relatively lower mobility (and thus on-current density. For a set of hold and boost voltages ($V_{hold}$, $V_{boost}$), where $V_{hold} \leq 0$ and $V_{boost} \geq V_{DD}$, lower $V_{th}$ yields higher on-state current (and thus access speed) but higher leakage (and therefore worse retention), and vice versa in a higher $V_{th}$. The $V_{th}$ can be modulated by the metal work function ($\Phi_m$), channel thickness (due to quantum confinement [30], and donor/acceptor trap density [31] and can be practically tuned using i). W-doping concentration ii). control of oxygen (O$_2$) flow during sputtering, or iii). post-fabrication annealing in nitrogen (N$_2$). Fig. 5a shows the IWO transistor's transfer curve within a tunable range using oxygen vacancy (donor) concentration, for which we test between a $V_{th}$ of -0.4 to 0.7 V. To facilitate the design space exploration of oxygen vacancy concentration and device width in an IWO 2T-GC, we developed ML-assisted compact models from simulation data captured in Sentaurus TCAD. The TCAD model is first calibrated to an experimentally demonstrated long-channel DG IWO transistor [11], then geometrically scaled into a 15 nm gate length ($L_g$) DG device and CAA device based on dimensions presented in [14]. In Fig. 5b-c, we explore the tradeoff between access time and retention time while tuning $V_{hold}$, $V_{boost}$, and $V_{th}$ in the write access transistor, and in Fig. 5d., we sweep access time and retention against the system performance of a 7 nm 128 MB DG-IWO 2T-GC LLC on PARSEC benchmarks (Section V). We find that beyond ~0.1 ms retention time, improvements in runtime strongly level off and that access time has a far more pronounced effect on performance. However, it should be considered that the impact of refresh scales with cache capacity (as the cumulative refresh time spent is a function of the number of rows per mat, here $N_{row}$ = 128). Additionally, as observed in JEDEC standards for commodity DRAM, refresh intervals are decided by tail bits in the <10$^{-3}$ percentile, indicating that the mean should be sufficiently high if the cumulative distribution function (CDF) is wide.

Furthermore, the read transistor must be able to move charges from the RBL quickly when a "1" is stored at the RA gate, which is challenging given that RBL parasitic capacitance is compounded when using AOS devices. Therefore, we suggest using an asymmetrically optimized 2T-GC with a higher dopant concentration in the read transistor (lower $V_{th}$) and a lower dopant concentration in the write transistor (higher $V_{th}$). Ideally, this read is carried out using logic-compatible voltages to prevent the use of bulky I/O level shifters, a reoccurring issue presented in prior studies of AOS 2T-GC, which would severely decrease the density and energy efficiency of AOS 2T-GC caches. Take, for example, a statically constructed 128 MB bank with the same organization as the top die in AMD's V-cache (Section 5). Without any level shifters, the macro can fit into a ~12 mm$^2$ frame; however, using level shifters on both the read and write access points (WWL, RWL) bumps this to ~28 mm$^2$, a >2.3× difference in total cache density.

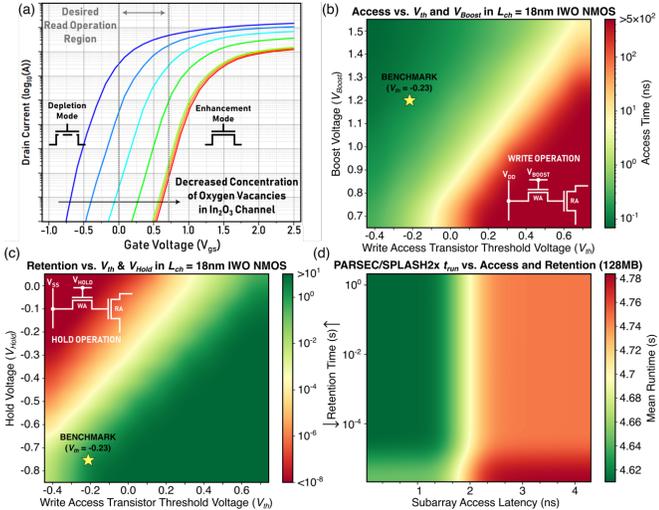

Fig. 5. (a) Relationship between threshold voltage ($V_{th}$), subthreshold swing (SS), and on-state current ($I_{on}$) with concentration of oxygen vacancies ($N_D$) in IWO transistors. Desirably, a narrow operation region is utilized to circumvent the extensive use of level shifters that impose significant constraints on density. (b)-(c) Access time vs. retention tradeoff in AOS 2T-GC. (d) Access and retention time vs. average runtime on PARSEC benchmarks, demonstrating a more substantial influence from access time on LLC performance. The data illustrated is interpolated bicubically.

## IV. NS-Cache and System Design

### A. Integration of Destiny and NeuroSim V1.4 as NS-Cache

To develop NS-Cache for early cache design exploration in advanced technology nodes, we carry over the organizational/circuit calibration methodology developed in the DESTINY/NVSim, the results of which are validated on RRAM and eDRAM prototype chips. To this end, we integrate advanced FinFET and nanosheet transistor predictive logic technology models validated using major foundry experimental data and IRDS projections [19] calibrated using Sentaurus TCAD device models. Layout optimization of standard cells in non-planar CMOS devices (e.g., folding, PN & dielectric wall separation, backside power rail) and advanced interconnect RC analytical models (effective copper resistance, FS-MS model) are upgraded through integration with NeuroSim V1.4. Specific technology, design methodology, interconnect parameters, and standard cell layout reduction techniques are discussed in detail by Lee et al. [18]. HD SRAM models for 14-1 nm nodes are carried over from NeuroSim through cell configuration files. DESTINY configuration files are backward compatible with NS-Cache, and a simple substitution of the tech node parameter can be used to test old designs in new technologies. Signal TSV analytical models are upgraded to estimate depletion capacitance using non-static depletion width derived by solving the 1D Poisson equation in cylindrical coordinates [32]. Scaled TSV and MIV [29] dimensions are added to support H3D and M3D integrated cache designs in scaled technologies. Circuit modeling for 2T-GC caches and full BEOL memory placement are added, details of which are the subject of the following subsections. Circuit models from NeuroSim are compatible with NS-Cache, allowing exploration of domains such as processing near memory within the proximity of ultra-large on-chip caches.

### B. M3D AOS 2T-GC Bank Modeling

Fig. 3 depicts the mat-level organization of an M3D AOS-based 2T-GC array and peripherals. NS-Cache is used to model the PPA of ❸-❿. A 2T-GC mat separates read and writes row decoders fed simultaneously by the subarray pre-decoder, used to activate RWLs and WWLs (❻-❼). Unselected cell leakage onto the RBL during a data "1" access degrades the sense margin and limits the potential mat size of 2T-GC [24]. To mitigate unselected cell leakage onto RBL and enable large mat size for maximized density, we consider the insertion of tri-state buffers in the final stage of each multistage RWL decoder driver to float unselected word lines at $V_{DD}$. In the write data path, level shifters (❽) maintain negative hold voltage $V_{hold}$ driven onto unselected lines and generate write pulses at $V_{boost}$ on selected lines in the write data path. Level shifter output drivers and precharger/write drivers (⓫) are each optimized for latency (limited to 10 stages) to reduce latency caused by higher xWL/xBL cumulative capacitance. Peripheries are connected over MIV arrays (❾) to BEOL memory tiers. The modeled memory array (❿) utilizes a folded-bit line architecture with adjacent partitioned memory arrays connected by RBLs to latching sense amplifiers (SAs) (⓬). A folded-bit line structure with dual-sided prechargers/write-drivers is adopted for stronger immunity to common-mode noise [33] and reduced RBL/WBL parasitic capacitance compounded by the switch to AOS devices. Each memory array partition includes a reference row activated during the adjacent array readout. Although 2T-GCs have a non-destructive readout, we maintain the use of fully connected sense amplifiers seen in 1T1C eDRAMs to reduce the latency that would otherwise be incurred by RBL and SA multiplexers. Several strategies limit the potential mat integration density, including adopting a folded BL architecture, dual-sided prechargers, latency-optimized buffering, and un-multiplexed blockwide SAs. The placement of BEOL memories allows these dedicated FEOL peripherals in favor of performance per unit area.

### C. FEOL and BEOL Partitioning

Memory array placement is performed over peripherals at the mat level to maintain a low interconnect overhead. The reduction in mat area propagates throughout the hierarchy by reducing the grid size of mats/subarrays and routing length, reducing IR drop, repeater insertion, and signal latency. Fig. 6 visualizes the partitioning and parasitic modeling strategies employed in our study of BEOL memory subarrays. First, a classical 2D mat is modeled to establish wireline parasitics and the footprint of peripherals and memory arrays, respectively. The memory array is then elevated to the BEOL and folded lateral to the largest planar dimension to fit the 2D footprint within the frame of the FEOL periphery. Based on an Mb/cost analysis performed in [11], the return on investment (ROI) of BEOL device stacking saturates beyond 7-8 M3D device tiers (due to the cost of photolithography). For this reason, we limit the number of stacked memory tiers to 4 layers, corresponding to 8 device tiers (2 tiers/2T-GC) or two algorithmic folding interactions to maximize the ROI of M3D integration. Mats requiring >2 folding iterations to optimize area are not considered during optimization. After folding, the memory array aspect ratio (AR) is used to reshape the FEOL tier layout to minimize the mismatch between FEOL/BEOL placement since xWL/xBL extensions in each dimension are needed to connect to MIVs placed at the FEOL boundary/edge. Stacked WLs and BLs are driven in parallel during operation to maximize the efficiency of folded memories, as the RC delay of each line is reduced in the folding process. MIV parasitics are multiplied by the maximum height (along several tiers) and are added to xWL and xBL RC alongside wireline extension parasitics before performing latency and power analysis.

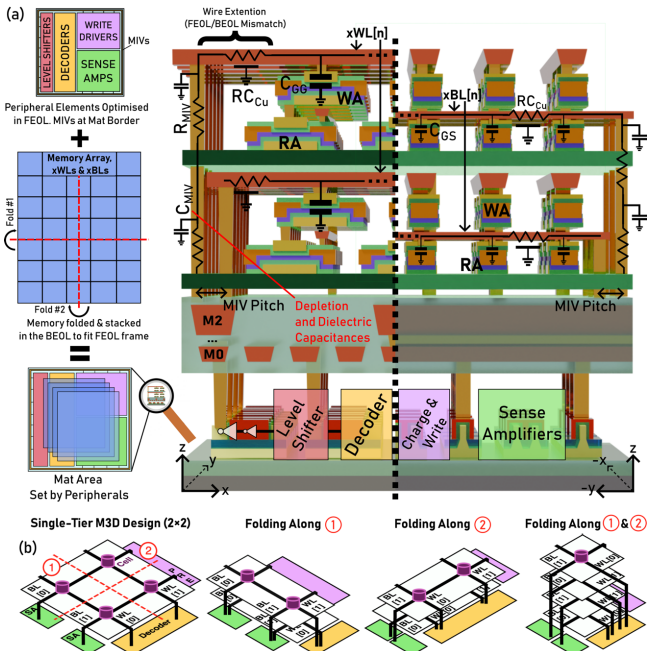

Fig. 6. (a) Spatial integration procedure and parasitics accounted for in M3D LLC at the mat level. Peripheral circuits (FEOL) and memories (BEOL) are partitioned and placed, and BEOL is folded into stacks to fit the FEOL frame. MIVs connect peripheral circuits to the memory array. (b) Peripheral allocation and xWL/xBL allotment in different tiers in a multi-tier folding scheme, using a simple 2×2 cell grid as an example.

### D. On Scaling Device Parasitic Capacitance

In AOS transistors, the overlap of source/drain regions and the gate contact over the oxide channel is critical in facilitating carrier injection driven by electron tunneling over the Schottky barrier. However, a byproduct of this overlap is large parasitic contact capacitances that impede performance in dense arrays, mainly due to RC latency when charging lines on which multiple cells are attached, but less obviously increasing the leakage within peripheral circuits due to the loading cost of buffers. Given a buffer chain with minimum input capacitance $C_{in}$ with $N$ stages, a cumulative effective fanout $F$, and a load of $C_L$, the optimal $N$ and size of inverter stage $k$ when optimizing for latency is given by:

$$s_k = C_{in}\left(\frac{c_L}{c_{in}}\right)^{\frac{k-1}{N-1}}, \quad k = [1, \dots, N] \quad (2)$$

$$\frac{\partial t_p}{\partial N} = \gamma + \sqrt[N]{F} - \frac{\sqrt[N]{F}\ln(F)}{N}, \quad F = \frac{c_L}{c_{in}}, \quad \gamma = C_{int}/C_g \quad (3)$$

$$N_{\gamma=0} = \ln(F) \quad (4)$$

The proportionality factor $\gamma$ is held at 0 in Equation 3 to obtain the closed-form solution that minimizes propagation time ($t_p$) given in Equation 4, thereby ignoring self-loading ($C_{int}$) [44]. Except for the minimum-sized initial stage, the transistors' width increases with $C_L$. A consequence is leakage power in each inverter scales with both $N$ and $s$, thus resulting in higher static power (and hence worse energy efficiency) in AOS 2T-GC arrays compared to their 1T1C eDRAM counterpart. In Fig. 7a, we project, using a 128 kB subarray, what the scaling benefits would be if $C_{gs}$ and $C_{gd}$ could be scaled down to FEOL Si transistor levels. Notably, even when scaled to Si-equivalent parasitics, 2T-GC arrays maintain worse leakage than 1T1C eDRAM due to the split R/W paths (and thus increased peripheral/buffer count). Furthermore, the eDRAM developed for IBM's Power processors utilizes silicon-on-insulator (SOI) technology; the buried oxide layer reduces the S/D capacitance by ~50-55% [34]. One avenue to mitigate this capacitance in AOS transistors is to minimize the overlap length of the S/D contacts; however, this reduces the current density and increases contact resistance. This tradeoff is illustrated in Fig.

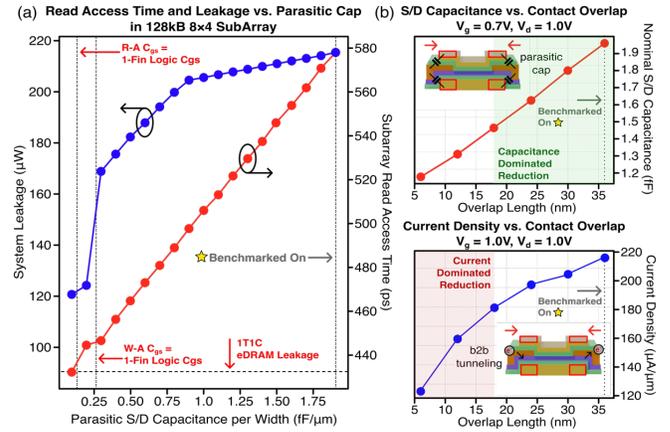

Fig. 7. (a) Effects of scaling down parasitic capacitance and translationally leakage (from buffers) and read timing in an AOS 2T-GC subarray ($L_g = 15$ nm, $L_{ov} = 36$ nm). (b) Illustration of the overlap length ($L_{ov}$) tradeoff between parasitic capacitance and on-state current density in an IWO device.

7b using TCAD, indicating a transition between capacitance-dominated and current-dominated reduction regimes.

### E. Tags Under Data Array and Shared Routing

Prior works on die-stacked DRAM caches have pointed out that storing tags in SRAM, although beneficial from a performance perspective, is not scalable due to the large footprint of SRAM tag banks whose capacity would grow linearly with data capacity (Fig. 8a) [35]. The same benefits and challenges posited by these works apply to on-chip alternative memory caches as well: SRAM cache latency is limited by interconnect latency, not device latency; thus, lower capacity macros (such as those used for tags) can deliver higher bandwidth and reduce cumulative miss time, but at the cost of exorbitant dedicated space on silicon when scaled for ultra-large caches towards the GB scale. However, in a monolithic 3D integrated LLC, the silicon space occupied by tag memories (specifically, the BEOL above them) may pose an opportunity

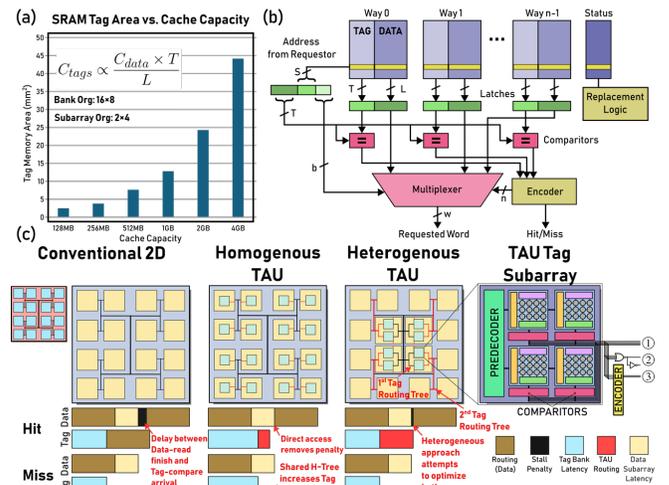

Fig. 8. (a) Si footprint of LLC (HP) SRAM tags scaled with data capacity at a fixed organization. (b) Architectural overview of the interplay between tag and data arrays within a $n$-way set associative cache, where $T$ is the tag bits, $L$ is the line size, $w$ is the word size, $S$ is the number of line-tag pairs per bank, and $n$ is the associativity. (c) Tag Arrays Under data (TAU) strategies, timing, and subarray modifications (direct $N_{DW}$) in ②-③.

## TABLE II
### ROUTING BUS-WIDTHS (GDL) IN TAU ARCHITECTURE

| Mode / Wire Type | | Address Wire ($N_{AW}$) | Broadcast Data Wire ($N_{BW}$) | Distributed Data Wire ($N_{DW}$) |
|---|---|---|---|---|
| **Normal Access** | Data | $\log_2(N_{Block}/A)$ | $\log_2(A)$ | $W_{Block,Data}$ |
| | Tag | $\log_2(N_{Block}/A)$ | $W_{Block,Tag}$ | $A$ |
| | TAU | $\log_2(N_{Block}/A)$ | $W_{Block,Tag}$ | $W_{Block,Tag}+2$ |
| **Sequential Access** | Data | $\log_2(N_{Block})$ | 0 | $W_{Block,Data}$ |
| | Tag | $\log_2(N_{Block})$ | $W_{Block,Tag}$ | $A$ |
| | TAU | $\log_2(N_{Block})$ | $W_{Block,Tag}$ | $W_{Block,Tag}+2$ |
| **Fast Access** | Data | $\log_2(N_{Block}/A)$ | 0 | $W_{Block,Data} \times A$ |
| | Tag | $\log_2(N_{Block}/A)$ | $A$ | $A$ |
| | TAU | $\log_2(N_{Block}^2/A)$ | $W_{Block,Tag}$ | $(W_{Block,Data} \times A)+A$ |

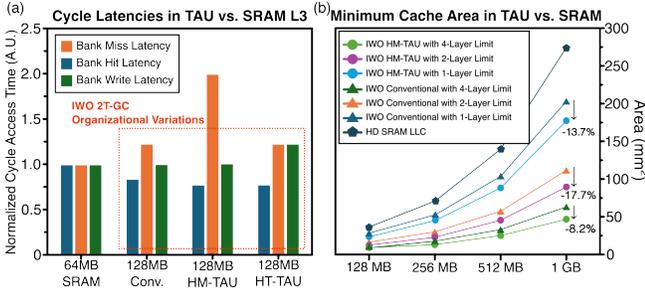

Fig. 9. (a) Comparison of cycle access latencies normalized to SRAM. (b) Minimum cache area of an LLC with fixed bank organization ($32\times16$) using M3D integrated AOS 2T-GC with/without HM-TAU and SRAM.

for tighter integration (and thus better utilization & shorter communication delays). We consider a scenario in which data memories in the BEOL are fabricated above SRAM tag subarrays/mats to develop ultra-high-capacity 2T-GC caches with SRAM tags that can share silicon real estate on the chip.

To understand why integrating tags and data banks can be advantageous, it is helpful to examine how tag and data memories interact in a cache (Fig. 8b). When a cache receives an address, the tag bits are sent to the tag memory to check for the presence of a valid entry; meanwhile, the line address is passed to the data memory to retrieve the corresponding data. In a sequential access scheme, which is more energy-efficient but incurs longer latency, the hit/miss confirmation arrives from the tag array before the data array is accessed on a hit or before the request is sent to the main memory on a miss. In contrast, in a "normal" access scheme (vernacular defined by NVSim), the line address and tag bits are supplied to the data and tag arrays simultaneously; however, the retrieved data remains in the data row buffer until the tag array confirms whether the transaction is a hit, after which the entry is flushed to distributed data lines ($N_{DW}$). Normal and sequential access schemes require inter-bank communication between the tags and data during a hit, which can be sped up through the proximity of outputs from tag comparators. In a strategy we call **T**ag **A**rrays **U**nder data (TAU), SRAM tag subarrays and mats are monolithically co-integrated using the same algorithm applied to AOS-based 2T-GC mat optimization. Using TAU, read bandwidth in AOS 2T-GC caches can be increased by ~20-25% over the conventional baseline using spatially separated tags and data memories.

The high-level details of our two TAU strategies are described in Fig. 8c, and Table II adds entries for changes to the global data line (GDL) width when employing TAU in each access mode modeled in NS-Cache. In a homogenous approach

(HM-TAU), the mat and subarray organization of tag and data banks are identical, and tag mats are directly mapped under data with identical partitioning of blocks. This approach benefits from tag subarrays having direct access to signal the data mats with trivial overhead during a cache hit. However, in banks with high mats/subarray, tag access latency/miss penalty is increased due to longer data routing length, making this more practical in sliced caches. High SRAM tag memory bandwidth is maintained since the latency at the tag subarray level is unperturbed. Heterogeneous TAU (HT-TAU) allocates a higher ratio of tag subarrays/mats under central data subarrays with the highest proximity to control/IO logic. A two-stage forward routing H-Tree is placed from control to TAU subarrays and TAU subarrays to outer subarrays (Fig. 8c). The miss penalty degradation seen in high subarrays/bank HM-TAU is overcome by bounding tags closer to control logic and splitting up tag routing. However, the larger FEOL footprint of HT-TAU subarrays increases the data access penalty due to longer local interconnects, increasing the write penalty. These changes to access time in a TAU-integrated IWO 2T-GC LLC are quantitatively assessed using a 128 MB macro in Fig. 9a.

Fig. 8c additionally highlights the key differences in a TAU tag subarray. In the traditional approach, a one-hot encoded string of bits (①) of width $A$ is routed from comparators back through control circuitry, where it is then encoded and passed to the data array through broadcast data wires ($N_{BW}$). A reduced set of bits in the TAU cache mat is sent back through the H-tree to indicate a miss to the requester using a one-hot encoded signal (②). Adding an onboard encoder (③) plays a crucial role in truncating outputs and driving them directly to the adjacent data bank periphery for processing during a hit.

Although the prospect of fabricating an arbitrary number of $M$-stacked AOS device tiers in the BEOL is promising from a footprint minimization standpoint, certain limitations remain. Photolithography is the costliest step in semiconductor manufacturing, and the number of lithography steps/masks grows linearly with $M$; consequently, the return on investment (ROI), measured by Mb/cost, drops significantly once the $M > 8$ [11]. Furthermore, demonstrations of M3D-stacked cells indicate that stacking can affect performance and tighten variation in lower tiers, potentially due to the formation of additional (undesired) defects [14]. Given the extra space for memory afforded by TAU, the minimum area of a cache with a fixed bank organization and reduced allotment of tiers (to mitigate challenges stated above) can be significantly reduced by >9× over SRAM towards GB-scale cache designs (Fig. 9b). This benefit increases with capacity but decreases with the number of allotted tiers.

## V. SYSTEM SIMULATION AND BENCHMARKING SETUP

### A. Quantization and Plugin into Gem5

To observe how single operation performance and energy affect high-performance computing system power and performance and to study the effects of architectural changes on cache performance, we interface NS-Cache with Gem5's Ruby protocol management system. The modifications made to Gem5 are meant to model the timing-accurate performance of refresh events. System-level latency metrics (e.g., hit, miss, and write latency) and subdivision metrics (e.g., subarray and routing latency) are quantized into discrete cycles using a parameterized clock frequency passed from NS-Cache to Gem5. To model the performance effects of refresh latency, frequency, and strategy, we introduce changes to Ruby's memory structures and protocols written in Gem5's SLICC language (Fig. 10). Based on the access mode (i.e., sequential, normal, fast), we add parameterized parallel (e.g., tag broadcast latency) and sequential (e.g., tag access latency) cycle penalties to characterize communication delays between tags and data memories. These changes better articulate bandwidth changes between cache designs to allow the study of alternative access modes.

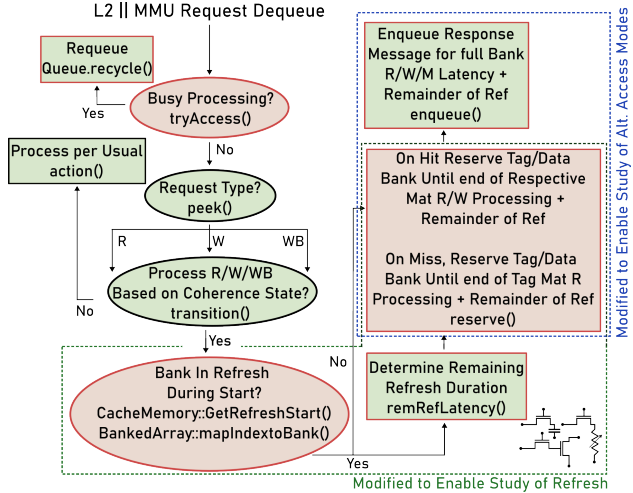

Fig. 10. Latency modeling refresh and access mode operation flow for data and tag banks in Gem5. Orange indicates Ruby structures and green for SLICC protocol definition in LLC. Modifications are outlined in red.

### B. Benchmarking Parameters and Baseline Comparison

To generate a qualitative comparison of IWO 2T-GC caches with caches constructed from other alternative memories, it is critical to evaluate against a competitive baseline at the same technology node. Unfortunately, as Section 2 highlights, STT-MRAM and 1T1C eDRAM have not been demonstrated experimentally beyond the 14 nm node. However, it is reasonable to expect that future scaling progress could be made in these devices if major foundries desire it. Thus, projected baseline parameters, shown in Table III, are generated at the 7 nm node using state-of-the-art reported parameters, design rules, and predictive technology parameters. STT-MRAM is modeled from a cache prototype reported by IBM in 14 nm technology [22], assuming the same MTJ is used with an access transistor in 7 nm design rules under the same MTJ current density requirements. The 1T1C eDRAM retention referred to in Intel's report [8] and the area and storage capacitance used in reported IBM POWER eDRAM [7] are scaled based on generational trends from GlobalFoundries' 22 nm to 14 nm eDRAM technologies (which is in partnership with IBM for manufacturing).

### C. Benchmarking Parameters and Baseline Comparison

AMD's Zen3 V-Cache integrates two stacked cache memory chiplets using hybrid bonding and TSVs [36]. The upper tier of this shared L3 cache comprises 64 MB of HD

### TABLE III
### DEVICE PARAMETERS USED IN NS-CACHE MODELING

| Device | Tech Node | Parameter | Value |
|---|---|---|---|
| IWO 2T-GC | 7 nm (DG) | Cell Area | $0.02052\ \mu m^2$ |
| | | Retention | 315 ms |
| | | Write Latency | 122 ps |
| | | $V_{Boost}$ | 1.2 V |
| | | $V_{Hold}$ | -0.75 V |
| | 3 nm (CAA) | Cell Area | $0.013\ \mu m^2$ |
| | | Retention | 251 ms |
| | | Write Latency | 421 ps |
| | | $V_{Boost}$ | 1.3 V |
| | | $V_{Hold}$ | -0.5 V |
| STT-MRAM | 7 nm | Cell Area | $0.0138\ \mu m^2$ |
| | | $R_{on}/R_{off}$ | 7970/19210 |
| | | Write Pulse Width | 4 ns |
| 1T-1C eDRAM | 7 nm | Cell Area | $0.0116\ \mu m^2$ |
| | | Retention | 201 µs |
| | | SN Capacitance | 5.4 fF |
| HD SRAM | 7 nm | Cell Area | $0.0276\ \mu m^2$ |
| | | PU:PD:PG Ratio | 1:1:1 (Fins) |
| | | Read Voltage | 0.7 V |
| | 3 nm | Cell Area | $0.0199\ \mu m^2$ |
| | | PU:PD:PG Ratio | 1:1:1 (Fins) |
| | | Read Voltage | 0.7 V |

### TABLE IV
### BENCHMARKING PARAMETERS USED IN GEM5 SE SIM

| Benchmark | Suite | Parameters |
|---|---|---|
| Particle Filter (PFil) | Rodinia | $dimX, dimY$: 1024, $N_{fr}$: 20, $N_{particle}$: $10^5$ |
| Needleman-Wunch (NW) | Rodinia | $maxRow/Col$: 4096, $penalty$: 10, $N_{th}$: 4 |
| Heartwall (HW) | Rodinia | $file$: test.avi, $N_{frame}$: 25, $N_{th}$: 4 |
| Backpropagation (BP) | Rodinia | $N_{elements}$: $10^7$ |
| LU Decomposition (LUD) | Rodinia | $size$: 12000, $N_{th}$: 4 |
| Pathfinder (PFin) | Rodinia | $width$: $10^5$, $N_{step}$: 250 |
| Volrend (VR) | Parsec/Splash | $file$: head_scaleandrew; $N_{rotate}$: $10^3$ |
| Ocean CP (OCP) | Parsec/Splash | $N$: 258, $tolerance$: $10^{-7}$, $D$: $2\times10^4$, $N_{step}$: $2.88\times10^4$ |
| Cholesky (CS) | Parsec/Splash | $postpass$: 32, $file$: d750.O |
| Fast-Fourier Transform (FFT) | Parsec/Splash | $m$: 26, $N_{processor}$: 4 |

### TABLE V
### SYSTEM PARAMETERS USED IN GEM5 SE SIM

| Parameter | Configuration |
|---|---|
| Memory | 16GB DDR4_2400_8x8 |
| CPU Configuration | 4 Core X86 Out of Order CPU |
| Clock Frequency | 3 GHz |
| Cache Line Size | 64 Bytes |
| L1 Configuration | 8-way 32kB Data, 32kB Instruction Cache per Core |
| L2 Configuration | 8-way 512kB Data Cache per Core |
| L3 Configuration | 16-way 64-128MB Data Cache |
| Replacement Policy | LRURP |

SRAM built using TSMC's 7 nm FinFET node that is partitioned into eight parallelly operating slices, built atop an 8-slice 32 MB layer of HD SRAM (Fig. 11). A ring bus in the bottom tier of the L3 cache carries transactions from cores (CCX) to the mapped slice, either in the bottom tier directly off the ring bus or in the upper tier through TSVs that map a pair of slices into a rank. To challenge the PPA of V-Cache without die-stacking, we first consider an iso-area/organization comparison of alternative LLC memories and HD-SRAM constrained based on the minimum-sized bit cell used in the upper tier of the V-Cache. This comparison is made to elucidate baseline device strengths and weaknesses. Topological images of 128 kB subarrays within the 2nd tier of V-Cache from [6] were analyzed to extract the dimensions and organization of mats, which we found to be 4×8 mats/subarray. Since the V-Cache uses DECTED ECC encoding [36], we estimate the number of total bits, assuming that ECC bits are uniformly distributed in the top and bottom tiers (17 ECC bits per 64 data bits). Using the area efficiency from a replicated subarray in NS-Cache using a minimum reported cell by TSMC (0.027 $\mu m^2$), we take the cell area to be the subarray area divided by the subarray capacity times the area efficiency, which yields an estimated cell size of ~0.0276 $\mu m^2$.

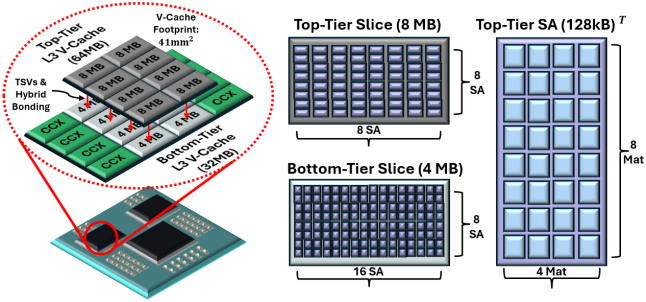

Fig. 11. Organization of AMD V-Cache high-level structure and bonding, slices, banks, and subarrays, divided into top and bottom tiers. Nomenclature reflects macros used in NS-Cache modeling.

In this first iso-area/organization comparison, we do not consider the slicing/ring bus attributes, instead opting to design a bank with 32×16 subarrays/bank and $N_{ASR} = N_{ASC} = 1$. Based on a model of an HD-SRAM cache with V-Cache top die organization (32×16, 4×8), we yield a footprint of 19.32 mm² for a 64 MB macro, which we set as a ceiling for the iso-area/organization comparison. Taking the same bank-level organization and alternative memory (Table III) caches optimized for read-write delay product (by optimization of mat organization), an STT-MRAM cache can fit double the capacity (128 MB) into the same footprint (18.78 mm²), and eDRAM scales this further to 15.51 mm². In a 2-tier design, a 128 MB IWO 2T-GC LLC fits into a mere 11.53 mm². In a 4-tier design, a 256 MB IWO 2T-GC requires 13.49 mm², yielding a 5.7× higher bit density (Mb/mm²) over SRAM and a 2.5× improvement over eDRAM. We use the 128 MB IWO 2T-GC cache for system benchmarking to maintain a fair apples-to-apples comparison of alternative memories. Later, in Section 6, we consider bandwidth extensions to maximize performance per footprint, then pit this LLC against a comprehensive model of the V-Cache.

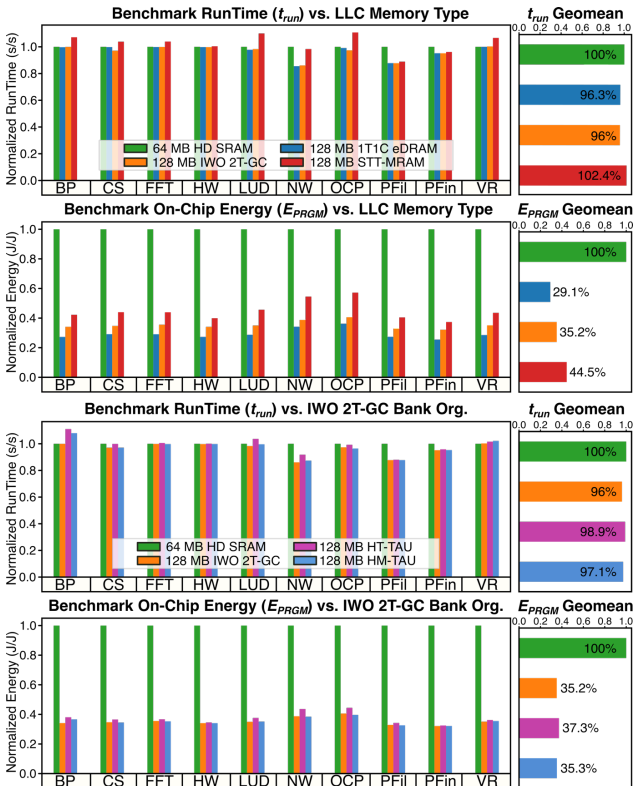

Fig. 12. (a) Runtime and on-chip energy comparison of LLC memories in an iso-area/organization benchmark. (b) Runtime and on-chip energy comparison of IWO 2T-GC integration strategies (conventional and TAU) in iso-area/organization benchmark. All results normalized to 64 MB SRAM macro (green).

### D. Benchmark Selection and Parameters

We use benchmarks from Rodinia [37] and PARSEC/Splash2x [38] suites with large problem sizes to better understand the effects of a refresh, bandwidth, and aggregate delay on performance on CPU-intensive workloads (Table IV). All benchmarks are compiled on OpenMP. The architectural parameters of our Gem5 simulation are detailed in Table V and are set to closely mirror a system built around AMD's Zen3 architecture. The Gem5 simulator [39] is fed cycle latency and refresh timing parameters from the bank and subarray level generated using NS-Cache.

## VI. BENCHMARKING RESULTS

### A. Iso-Area/Organization Comparison of 7 nm LLCs

Key benchmark results for runtime and on-chip LLC energy of each baseline 7 nm cache macro are shown in Fig. 12a. In most applications, 128 MB STT-MRAM lags in multi-core performance compared to other cache memories with 2.4% higher runtime than the SRAM geometric average. Exceptions to this rule are applications with a high data working set size, such as the arithmetic-heavy Needleman-Wunch, Particle Filter, and Pathfinder, where the increased hit rate (a result of the higher capacity) significantly reduces the total runtime. The 128 MB 1T1C eDRAM and 128 MB IWO 2T-GC LLC macros consistently deliver higher performance than STT-MRAM (~6.4% geomean) due to lower bank access latency and higher bandwidth (because of lower reservation latency at the subarray level) and perform strongly against the SRAM baseline (~3.7-4% geomean) thanks to increased capacity and shorter bank-level retrieval time. However, the runtime changes between 1T1C eDRAM and IWO 2T-GC are more subtle. Though 1T1C eDRAM's frequent refresh hinders its cache availability, decreasing both read and write bandwidth, the access time (particularly during write operation) of the AOS alternative plays a comparable role in decreasing IWO 2T-GC cache performance ($WBW_{IWO} \approx 0.35 \times WBW_{1T1C}$), leading to comparable mean runtime between the two systems. The performance gains of 1T1C eDRAM and IWO 2T-GC highly depend on access pattern and runtime range. To illustrate, the 128 MB IWO 2T-GC macro is the performance frontrunner in programs such as OceanCP and Cholesky with a higher read-to-write ratio, where the long runtime of OceanCP (~19 s) and frequent access pattern may increase the impact of eDRAM's refresh periodicity cache performance. Alternatively, 1T1C eDRAM is the performance frontrunner in a benchmark such as LU-Decomposition with both heavy read and write accesses. In bursty high-traffic applications (bytes/FLOP) like Backpropagation and FFT, SRAM's high RBW/WBW and lack of refresh maintain strong runtime compared to eDRAM and IWO 2T-GC caches.

Using L3 cache and runtime statistics from Gem5, we calculate LLC energy consumption in NS-Cache by:

$$E_{PRGM} \approx N_H E_H + N_M E_M + N_W E_W + \frac{t_{run} N_{row} E_r}{t_r} + P_s t_{run} \quad (5)$$

$N_H$, $N_M$, and $N_W$ are the number of hits, misses, and writes (Gem5); $E_H$, $E_M$, $E_W$, and $E_R$ the energy per operation for hits, misses, writes, and line-refreshes; $t_{run}$ is the program runtime; $N_{row}$ is the number of rows in a mat, and $P_s$ is the static power consumption of the LLC. As seen in the geometric mean comparison in Fig. 12a, all three alternatives to SRAM can cut $E_{PRGM}$ by over 50% despite double capacity. A breakdown of the geomean power consumption of each macro is captured in Fig. 13a. The IWO 2T-GC energy consumption, much like SRAM, is strongly dominated by leakage (90%). However, unlike SRAM, the leakage dominance results from the increased peripheral count and multi-stage buffering used to handle the higher device parasitic capacitances, not the memory devices themselves. Because static energy consumption is higher in the periphery, techniques such as power-gating can reduce cache idle energy without losing the contents in memory when under low load. As discussed in Section IV, this leakage may be halved if the parasitic capacitance can be reduced to logic-comparable levels. Although dynamic refresh energy constitutes a large percentage of overall eDRAM energy consumption, it is evident from Fig. 12a & 13a that the high write dynamic energy in STT-MRAM and leakage stemming from large current sense amplifiers (CSA) used to sense the small TMR is far more burdensome. Roughly 13% of the 1T1C eDRAM macro energy is consumed by refresh in a macro using $N_{row}$=128. In contrast, a negligible (<<1%) amount is consumed in IWO 2T-GC macro despite having much higher energy consumption per refresh operation (split W/R paths + energy cost $\propto$ CV$^2$).

Benchmark results for power and performance comparison between TAU integration strategies and SRAM are shown in Fig. 12b. Both TAU organizational strategies maintain lower runtime than their SRAM counterpart (HM-TAU: 2.9%, HT-TAU: 1.1%) but perform worse in runtime than their conventional organization counterpart using spatially separated tag and data banks. However, in bursty high read-traffic applications, such as OceanCP and Cholesky, HM-TAU strategies can improve performance by improving hit data retrieval latency. Additionally, as is mentioned briefly in Section 4, if a cache were to employ slicing (therefore reducing the routing mismatch between tags and data), HM-TAU may prove increasingly viable due to a smaller shift in the miss penalty. Though HT-TAU offsets the longer miss penalty, the additional write penalty (2 cycles) exacerbates the already low write bandwidth in IWO 2T-GC caches. However, HT-TAU may provide a better alternative with negligible miss penalty degradation if one develops a BEOL-compatible memory with a more favorable write latency/lower device parasitics.

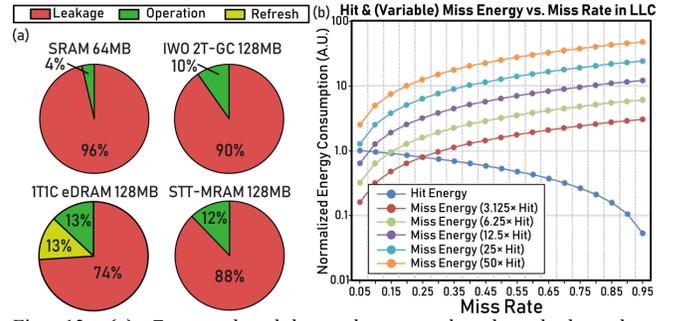

Fig. 13. (a) Energy breakdown between benchmarked caches. Refreshing in AOS caches has a minimal impact on total power consumption (~0.04%). (b) Variable hit-and-miss energy in LLC, where miss energy is a multiple of hit energy. Typical estimates for miss operation energy are ~50-200× that of a hit, illustrating the necessity of minimizing off-chip data movement with higher LLC density.

When discussing the impact of an LLC on the cumulative energy consumption, it is also essential to consider the cost of off-chip data movement (usually omitted in prior work), which routes through large I/O transistors and pads and across a PCB/interposer to main memory. According to some estimates, data movement between the processor and main memory constitutes >63% of total system energy consumption [40]. In the literature, the energy cost of a miss in the LLC is an estimated ~50-200× that of a hit [41], which, as shown in Fig. 13b, even with a near-perfect hit rate, the energy cost of misses strongly outpaces hit energy. Furthermore, the rate at which miss energy decreases is steeper as the miss rate decreases, further emphasizing the utility of high-capacity LLCs. Using CACTI-IO, we estimate that the miss energy in the 7 nm cache design when interfacing with a DDR4 DIMM is ~92× higher than a 64 MB SRAM cache miss. This estimation is based on a monolithic die interfacing with the main memory. However, in Zen3+ cores, CCD and I/O dies are partitioned, meaning on a miss, the CCD must first communicate over the interposer to the I/O die before the processed request is moved to DRAM, further penalizing misses and emphasizing further the need for high-capacity LLCs to offset the energy cost of data movement.

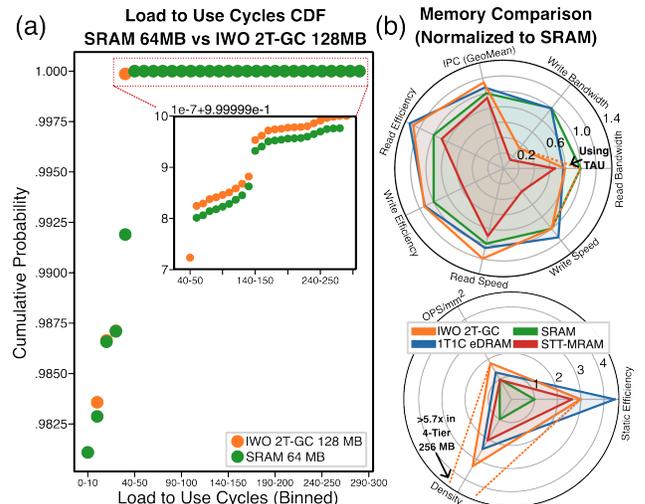

Fig. 14. (a) Load to use cycle CDF comparison of SRAM and IWO 2T-GC macros. (b) Figure of merit comparison between baseline memory macros.

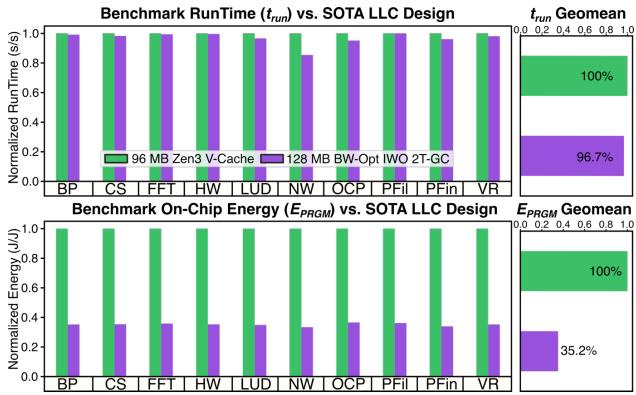

Fig. 15. Comparison of AMD V-Cache and BW-optimized IWO 2T-GC performance within constrained the same (planar) footprint.

| Metric | SRAM | CAA-IWO 2T-GC | % Change |
|---|---|---|---|
| Capacity | 128 MB | 128 MB | N/A |
| Area | 24.147 mm² | 10.545 mm² | -56% |
| Read Latency | 9.794 ns | 8.016 ns | -18% |
| Write Latency | 5.613 ns | 6.504 ns | +16% |
| Read Energy | 649.7 pJ | 461.6 pJ | -29% |
| Write Energy | 616.2 pJ | 654.8 pJ | +6% |
| Leakage | 370.66 mW | 209.949 mW | -43% |

To understand the driving changes seen at the cores leading to performance improvements in IWO 2T-GC caches, we plot the CDF of the load to use cycles (the number of cycles after a load instruction is executed in which the data is retrieved and begins processing) for the OceanCP benchmark in Fig. 14a for the SRAM 64 MB and IWO 2T-GC 128 MB caches. Though the vast majority of data is processed quickly (0-10 cycles) in both cases (given identical upper-level memory hierarchy), significantly more requests are processed in the 50-60 cycle range within the IWO 2T-GC cache than SRAM (~0.75%). This narrowing of the CDF at this range can be understood from the shorter mean retrieval time in the IWO 2T-GC cache due to shorter routing (and greater density) and increased hit rate as a function of greater capacity.

Fig. 14b summarizes the key figures of merit of different memories in the iso-area/organization comparisons. Benchmarking results demonstrate that 1T1C eDRAM implementations can deliver the best on-chip energy efficiency (specifically as a function of the better static efficiency measured in $W^{-1}$) among LLC memory candidates while delivering comparable write bandwidth to SRAM, only faltering on the read bandwidth (due to destructive read and refresh). Outside of the density and Ops/mm² benefit, STT-MRAM struggles to keep up with alternative memories and SRAM on energy and latency metrics. Although IWO 2T-GC caches have higher leakage than 1T1C eDRAM, they offer comparable read bandwidth with SRAM (using TAU) and faster read speed (shorter routing latency), giving them the most significant performance benefit over potential LLC memory candidates. However, the most profound benefit of IWO 2T-GC caches is in the density, which, despite requiring a bulky set of level shifters, can achieve >5.7× greater density than SRAM in a 4-tier 256 MB design. With the continued improvements in device characteristics (i.e., parasitic capacitance reduction), AOS 2T-GC's potential for improved write performance and leakage reduction (exemplified in Section 4) may further improve its viability as an LLC candidate.

### B. Bandwidth Optimized AOS Benchmarks vs. V-Cache

The iso-area/organization comparison demonstrates that IWO 2T-GC LLCs can deliver excellent performance, energy efficiency, and density when compared to the 64 MB SRAM baseline. To benchmark IWO 2T-GC memories against the state-of-the-art (SOTA, AMD's V-Cache), we use NS-Cache to generate an IWO 2T-GC cache that maximizes the performance within the allocated footprint (41 mm²). First, we reorganize the partitioning of the LLC (16 8MB slices, each 8×16 subarrays) to maximize hardware parallelism. We utilize HM-TAU to limit the macro size using two stacks of memories and improve read bandwidth for hits, given that slicing reduces the routing overhead within each partition. This M3D cache fits within a 25.41 mm² footprint, a 38% reduction over the planarized 2D footprint of V-Cache or a 69% reduction in total silicon footprint (accounting for hybrid-bonded die). To model the whole 96 MB of V-Cache, we use Ansys HFSS to accurately model the hybrid bonding pad and TSV interconnect parasitics based on AMD's reported bonding density and chip teardowns from [36], which are then used for power calculation of inter-tier communication. 8×16 subarray/slice 8 MB SRAM slices are modeled in NS-Cache using the upper tier cell size of V-cache to estimate the quantized latency parameters of each slice. An additional latency penalty is allocated i). for communication to the upper tier of 4 cycles based on [36], and ii). in each direction for the ring bus transmission that connects the slices in both the SRAM and IWO 2T-GC slices calculated assuming the ring bus is placed in upper metallization (M7-M8) with R = 0.05 Ω/μm and C = 0.2 fF/μm. Gem5 parameters for the level of parallelism in the data portion of the V-Cache are reflective of the accumulation of the 32×32 8-slice 32 MB bottom tier and 32×16 8-slice 64 MB upper tier. The resultant data access time of the L3 V-Cache model is ~10 ns, comparable to the AIDA64 estimate of ~12-13 ns produced by probing a Ryzen 5800X3D CPU [45]; however, this disparity is reasonable given that AIDA measurement-based estimations also account for digital controller and ECC overheads during data retrieval that isolated in the estimation produced for the LLC macro.

Fig. 15 depicts findings from the high-performance comparative study. Despite delivering over 2× the capacity in a single die, IWO 2T-GC consumes ~35.2% of the energy and reduces runtime by 3.3% over V-Cache. Write bandwidth-intensive applications such as Particle Filter and Heartwall continue to have marginal benefits over the SRAM implementation, given the lower write bandwidth in the IWO 2T-GC macro, leaving room for future improvement with innovation in device geometry and material selection.

### C. CAA AOS 2T-GCs Towards the 3nm Node

Since 2023, major foundries have been preparing for high-volume 3 nm process production. Therefore, alternative LLC candidates must be scalable to become viable on-chip SRAM substitutes. To this end, we compare 3 nm SRAM and high-density CAA-IWO 2T-GC LLCs using device characterization derived from our compact model analysis, which are compared in Table VI. To maintain consistency with the iso-area comparison performed in Section 7A, we model an SRAM and IWO 2T-GC 128 MB banks using a 32×16 subarray organization in NS-Cache and perform a direct analysis of PPA metrics at 3 nm technology using the minimum reported SRAM cell size [3]. CAA IWO 2T-GCs with vertically stacked write/read transistors enable further density scaling. In 3 nm, an IWO 2T-GC LLC macro delivers 29% lower read latency and 18% lower read energy with comparable write performance/energy at ~0.5× the leakage and total area compared to 3 nm SRAM. This is despite longer cell access latency in CAA structures compared to their DG counterparts (Table IV), a change consistent with the finding that interconnect latency plays a significant role in performance degradation in leading-edge SRAM. However, the longer cell latency attributed to the weaker *SS* (using a single gate) and the inherently larger overlap stemming from the modified MIM structure leads to a more significant proportion of the latency being spent at the subarray level, reducing the throughput of the cache.

## VIII. CONCLUSION

This paper presents a systematic study and optimization of ultra-large AOS 2T-GC LLCs to realize the potential of AOS memories as a high-performance SRAM substitute. A newly integrated cache modeling tool, NS-Cache, is developed for open-source and is interfaced directly with Gem5 to conduct system-level benchmarking on advanced technologies. W-doped $In_2O_3$ (IWO) 2T-GC cells are optimized with an asymmetric W-doping profile to achieve low latency and high retention at logic-compatible access and hold voltages. Optimized IWO 2T-GC caches achieve 14.4% higher multi-core performance, 1.8× Ops/mm², and >3× greater energy efficiency in a macro with twice the memory at 59.6% of the Si footprint when compared to a state-of-the-art SRAM macro implementation in 7 nm technology. TAU M3D integrated SRAM tags are presented as a solution to increase the viability of reaping SRAM tag bandwidth by reducing the total integration area by 17.7% and improving read bandwidth by ~20-25%. CAA structures are assessed to address the scaling of IWO 2T-GC eDRAM towards the 3 nm technology node, demonstrating a 56% smaller footprint, 18% quicker read access, and 43% lower leakage for a 128 MB macro. In summary, AOS 2T-GC memories demonstrate strong potential as a cache memory substitute in high-performance systems, offering greater multi-core performance and density than other alternative cache memory devices and reduced energy consumption over SRAM.